\documentclass[prd,showpcs,amsmath,amssymb,nofootinbib,longbibliography,twocolumn,showpacs,notitlepage,superscriptaddress]{revtex4-1}

\newcommand{\h}{\mathcal{H}}

\def\MM{M_{P}}
\newcommand\Lr{\rm L}

\usepackage{xcolor}

\usepackage[utf8]{inputenc}

\usepackage{latexsym}
\usepackage{epsfig}
\usepackage{graphicx,subfigure}
\usepackage[normalem]{ulem}
\usepackage{color}
\usepackage{xcolor}
\usepackage{multirow}
\usepackage{hyperref}
\usepackage{numprint}
\usepackage{eucal}
\usepackage{amsmath}
\usepackage{amssymb}

\newcommand{\geffs}{G_{\rm eff}^\Psi} 
\newcommand{\geffsh}{G_{\rm eff}^{\Psi+\Phi}}

	\usepackage{amsmath}
	\usepackage{amsfonts}
  	\usepackage{amssymb}
	\usepackage{makeidx}
	\usepackage{amsfonts}
	\usepackage[ansinew]{}
	\usepackage[usenames,dvipsnames]{pstricks}
	\usepackage{epsfig}

    \usepackage{graphicx}

	\usepackage{float}

\newcommand{\be}{\begin{equation}}
\newcommand{\ee}{\end{equation}}
\newcommand{\bea}{\begin{eqnarray}}
\newcommand{\eea}{\end{eqnarray}}
\newcommand{\ba}{\begin{align}}
\newcommand{\ea}{\end{align}}

\makeindex

\begin{document}

\title{Testing the consistency of gravitational waves and large scale structure constraints on dark energy}

\author{Antonio Enea Romano}
\affiliation{Instituto de Fisica,Universidad de Antioquia,A.A.1226, Medellin, Colombia}
\affiliation{ICRANet, Piazza della Repubblica 10, I--65122 Pescara, Italy}
\author{Juan Manuel Cardenas Mancipe}
\affiliation{Instituto de Fisica,Universidad de Antioquia,A.A.1226, Medellin, Colombia}

\begin{abstract}
Gravitational wave (GW) astronomy has opened a new window on the Universe, allowing to obtain constraints on dark energy and gravity independent from other electromagnetic waves observations, such as large scale structure (LSS).
For the purpose of investigating the consistency between  different observations the effective field theory (EFT) of dark energy is a useful tool, allowing to derive model and parametrization independent consistency relations (CR) between the effective gravitational constant, the slip parameter, the gravitational and electromagnetic luminosity (EM) distance, and the speed of GWs.
We test the constant brading and no-slip CRs,  by comparing for the first time the constraints on the effective gravitational coupling obtained from LSS observations with those from GW events with and without electromagnetic counterparts, confirming the  validity of the CRs at the current level of experimental uncertainty.  
The event GW170817 and its electromagnetic counterpart provides a constraint of the effective gravitational constant with an accuracy comparable with LSS constraints, while the analysis of GW events without electromagnetic counterpart are consistent,  but do not have a constraining power comparable to LSS observations. Beside allowing to test the consistency between independent observations, the CRs can be used to estimate the effective gravitational coupling with GWs at high redshift, where other observations are not available.
\end{abstract}


\maketitle

\textbf{Introduction---}
The theory of General Relativity (GR) \cite{Einstein:1915ca} has been tested with different types of astrophysical and cosmological observations and it provides the foundation of the standard cosmological model.
It has proven successful not only in explaining observations that Newton's theory of gravity \cite{Newton:1687eqk} could not account for, such as the precession of Mercury's perihelion, but also in predicting previously unobserved phenomena, such as gravitational lensing or gravitational waves \cite{LIGOScientific:2016aoc}. Despite these unquestionable successes, modified gravity theories (MGT) are under current active investigation, motivated by the goal to provide a fundamental explanation of dark energy.
In the framework of theoretical physics MGTs can be formulated in terms of an action which modifies the Einstein-Hilbert action, and the theoretical predictions can then be computed using cosmological perturbation theory or numerical simulations. Large scale structure observations allow to test the effects of MGTs on scalar perturbations, and can be conveniently understood in terms of an effective gravitational constant \cite{Ishak:2024jhs}, while GWs observations \cite{LIGOScientific:2016aoc} allow to study the effects on tensor perturbations.
Since both LSS and GWs observations are theoretically expected to be affected by the MGTs effects derived from the same action, it is expected that the MGTs constraints from these two sets of observations should be related. 
MGTs are normally studied assuming some phenomenological ansatzes, however this can sometime cause a misestimation of the  
observables \cite{Linder:2016wqw}.
It is therefore important to develop parametrization independent tests   relating directly physical observables. 
In this regard the effective field theory of dark energy \cite{Gubitosi:2012hu} is the ideal tool  to investigate the relation between the MGTs effects on different observables, allowing to derive consistency relations  \cite{Romano:2025pcs} between the effective gravitational constant, the slip parameter, the gravitational and electromagnetic waves (EMW) luminosity   distances, the speed of gravitational waves  and the sound speed. We apply the constant brading consistency relation to map GW-EMW distance ratio observational constraints to LSS effective gravitational constant constraints, showing that the two sets of observations are consistent, and allowing to compare what set of observations allows to obtain the most stringent constraints on the variation of the effective gravitational coupling on cosmological scales.

\textbf{GW effective action---}
The quadratic effective field theory action (EFT) of perturbations for a single scalar  dark energy field was derived in \cite{Gleyzes:2013ooa}.
The EFT action for tensor modes \cite{Gubitosi:2012hu,Gleyzes:2013ooa} is 
\be
S_{\gamma}^{(2)} =\int d^4 x \, a^3 \frac{\MM^2 f }{v^2_{\rm GW}} \left[   \dot{\gamma}_{ij}^2 -\frac{v^2_{\rm GW}}{a^2}(\partial_k \gamma_{ij})^2 \right]\,, \label{lgamma}
\ee
where the GWs speed is related to the EFT action coefficients by
\be
v_{\rm GW}^2 =  \left(1+\frac{2m_4^2}{\MM^2 f}\right)^{-1}\;,\label{vEFT}
\ee
where $f(t)$ sets the tensor-sector Planck mass, while $m_{4}(t)$ alters the tensor gradient term and thus the gravitational-wave speed $v_{\rm GW}$. An effective approach consistent with Eq.(\ref{lgamma}) and including higher order effects was developed in \cite{Romano:2022jeh,Romano:2024apw}, showing that the effective speed can acquire an additional frequency and polarization dependency, but in this paper we will only consider the leading order effects corresponding to Eq.(\ref{lgamma}).

\textbf{Effects of modified gravity on gravitational waves---}
In the literature of modified gravity the quantity $M^2_*=\MM^2 f /\,v_{\rm GW}^2$ is often introduced, in terms of which the action, using conformal time, takes the form
\be
S_{\gamma}^{(2)} =\int d^4 x \, a^2 M^2_* \left[   {\gamma'}_{ij}^2 -v^2_{\rm GW}(\partial_k \gamma_{ij})^2 \right]\,. \label{lgammaeta}
\ee
Note that $v_{\rm GW}$ depends on the ratio of two coefficients of the EFT action, $m_4$ and $f$, so that observational constraints on $v_{\rm GW}$ are  mapped into constraints of this ratio, not of the individual coefficients of the action. 
After defining $\Omega=M_p\sqrt{f}$ the equation of motion corresponding to the effective action is 
\be
\gamma_{ij}''+2  \h\Big(1-\frac{v_{\rm GW}'}{\h v_{\rm GW} }+\frac{\Omega'}{\h\Omega}\Big) \gamma_{ij}'-v^2_{\rm GW} \nabla^2 \gamma_{ij}=0  \,,\label{heft}
\ee
which can be solved with a WKB approximation \cite{Nishizawa:2017nef} on sub-horizon scales, allowing to derive the effect on the GW-EMW luminosity distance ratio $r_d$
\cite{Romano:2023ozy,Romano:2023xal}  

\be
r_d(z)=\frac{d^{\rm GW}_{\Lr}(z)}{d^{\rm EM}_{\Lr}(z)}=\sqrt{ \frac{f(0)v_{\rm GW}(z)}{f(z)v_{\rm GW}(0)}}=\\
\frac{M_*(0)}{M_*(z)} \sqrt{\frac{v_{\rm GW}(0)}{v_{\rm GW}(z)}}   \,,\label{rd}
\ee
where $z$ denotes the redshift.

\textbf{Effects of modified gravity on scalar perturbations---}
In the conformal Newtonian gauge the scalar perturbations of the expanding Universe metric take the form
\be
ds^2 = -\left( 1 + 2 \Psi \right) dt^2 + a^2(t) \left( 1 - 2 \Phi \right) 
d\vec{x}^2 \ , 
\ee
and the perturbed field equations can be written as \cite{Linder:2015rcz}
\bea 
& &   \nabla^{2} \Psi = 4\pi a^{2} G^{\Psi}_{\rm eff} \rho_m\,\delta_ m \,, \\  
& &   \nabla^{2} \Phi = 4\pi a^{2} G^{\Phi}_{\rm eff} \rho_m\,\delta_ m \,, \\
& &   \nabla^{2} (\Psi+\Phi) = 8\pi a^{2} G^{\Psi+\Phi}_{\rm eff} 
\rho_m\,\delta_m \,, 
\eea 
where the effective gravitational constant is \cite{Linder:2015rcz}
\be 
\frac{G^{\Phi}_{\rm eff}}{G_N}= \frac{2M_p^2}{M_\star^2} 
\frac{[\alpha_B(1+\alpha_T)+2\alpha_{MT}]+\alpha_B'}{(2-\alpha_B)[\alpha_B(1+\alpha_T)+2\alpha_{MT}]+2\alpha_B'} \,, \label{eq:geff}
\ee 
and the gravitational slip $\bar\eta$ is 
\be
\bar\eta=
\frac{(2+2\alpha_M)[\alpha_B(1+\alpha_T)+2\alpha_{MT}]+(2+2\alpha_T)\alpha_B'}{(2+\alpha_M)[\alpha_B(1+\alpha_T)+2\alpha_{MT}]+(2+\alpha_T)\alpha_B'} \,.\label{eq:etafull} 
\ee
where $\alpha_{M}$ denotes the Planck-mass running, $\alpha_{B}$ quantifies the scalar–metric kinetic mixing (braiding), and $\alpha_{T}$ characterizes the tensor-speed excess. We also have used the notation $\alpha_{MT}=\alpha_M-\alpha_T$, the above equations were obtained by applying the EFT to Horndeski theories \cite{Linder:2015rcz}, we denote with prime  $d/d\,\rm{ln} \,a$, with a the scale factor, and we use the definition of gravitational slip  $\bar\eta$ \cite{Bellini:2014fua}
\be 
\bar\eta=\frac{2\Psi}{\Psi+\Phi}=\frac{\geffs}{\geffsh} \label{bareta}
\,,
\ee
which is related to the other definition of slip $\eta$ by 
$\eta=\Psi/\Phi=G^{\Psi}_{\rm eff}/{G^{\Phi}_{\rm eff}}=\bar\eta/(2-\bar\eta)$.
In the $\alpha_i$ parametrization the speed of GWs is given by $
v^2_{\rm GW}=1+\alpha_T$ 
and the GR limit corresponds to $\{f=1,\alpha_i=0\}$, implying $G^{\Phi}_{\rm eff}/G_N=G^{\Psi}_{\rm eff}/G_N=\bar\eta=\eta=1$.

\textbf{Constant brading consistency relation---}
Since $G^{\Psi}_{\rm eff}$, $\bar\eta$ and the distance ratio $r_d$ depend on the same EFT coefficients and property functions, we can combine Eq.(\ref{eq:geff}) with Eq.(\ref{rd}) and Eq.(\ref{eq:etafull}) to obtain consistency relations between gravitational waves and large scale observations.
Assuming constant brading the following relation between observable quantities can be obtained \cite{Romano:2025pcs}

\be
8 \pi M_p^2\,  \frac{2-\bar\eta}{\bar\eta}G^{\Psi}_{\rm eff}=\frac{2}{2-\alpha_B}\left[\frac{d^{\rm GW}_{\Lr}(z)}{  d^{\rm EM}_{\Lr}(z)}\right]^2 \frac{v_{\rm GW}(z)}{v_{\rm GW}(0)}\,, \label{GeffCB}
\ee
where we have assumed $f(0)=1$ to account for local constraints, i.e. $G^{\Psi}_{\rm eff}(0)=G_N=1/8 \pi M_p^2$, and   $z$ is the redshift. 
The above equation establishes a simple  and intuitive relation between different observables which could be affected by gravity modification:  the effective gravitational coupling, the slip, the electromagnetic and gravitational luminosity distances,  and the speed of gravitational waves. The l.h.s. involves large scale structure observations, while the r.h.s. is related to gravitational waves observations. Alternatively it can be considered a consistency relation between the effects of modified gravity on scalar and tensor perturbations.

In the the luminal no-slip no-brading limit the CR in Eq.(\ref{GeffCB}) is in agreement with no-slip gravity \cite{Linder:2018jil}, and  with some non local theories \cite{Belgacem:2017ihm}. 
Note that the CR is also satisfied by GR, since in this case $v_{\rm GW}=1$, $G_N=1/8 \pi M_p^2$, $\alpha_B=0$ and $d^{\rm GW}_{\Lr}=d^{\rm GW}_{\Lr}$. This is expected, since GR is just another constant brading theory which can be formulated in the EFT framework. In the following sections we will take advantage of the parametrization independency of the CR to test the consistency of constraints on MGTs obtained analyzing different sets of  observational data.

\textbf{Comparing LSS and GWs estimations of $G^{\Psi}_{\rm eff}$---}
Large scale structure observations can be used to constrain $G^{\Psi}_{\rm eff}$, and the recent DESI \cite{Ishak:2024jhs} results are setting  stringent constraints on its redshift dependence.
Assuming the GW speed to be the same as the speed of light, the consistency relation gives a relation between $G^{\Psi}_{\rm eff}$ and the GW-EMW distance ratio.

\begin{figure}
\includegraphics[width=8cm]{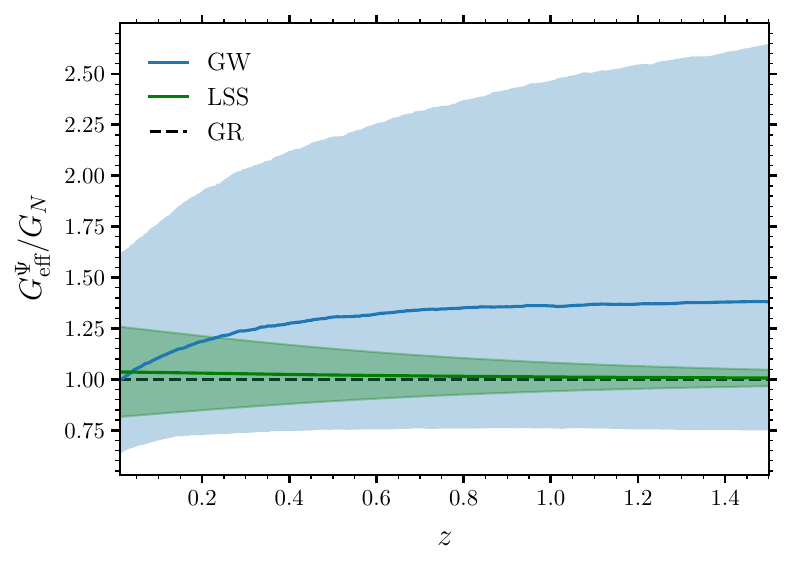}
\caption{The effective gravitational coupling $G_{\text{eff}}^{\Psi}/G_N$ is plotted as a function of redshift.
The posterior obtained from the analysis of LSS data \cite{Ishak:2024jhs} is plotted in green.
The posterior obtained from GWs observations using eq.(\ref{GeffCB0}) is plotted in blue, assuming no-brading, i.e. setting $\alpha_B=0$, and using the posterior for $\eta$ obtained from LSS data analysis \cite{Ishak:2024jhs}. This plot shows the results obtained adopting  the $\{\Xi_0,n\}$ parametrization for the GW-EMW distance ratio. The solid lines correspond to the median, while the shaded bands indicate the $1\sigma$ confidence interval of the posterior.}
\label{GeffChiDESI}
 \end{figure}

\begin{figure}
\includegraphics[width=8cm]{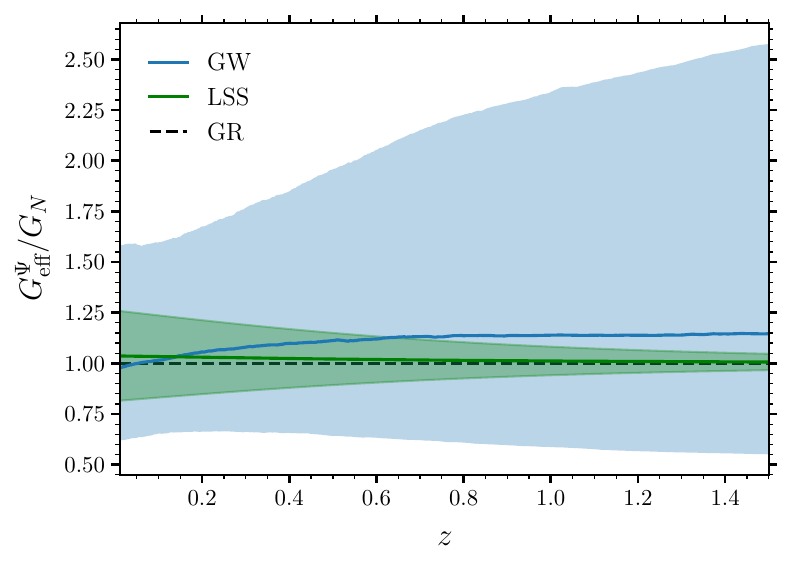}
\caption{The effective gravitational coupling $G_{\text{eff}}^{\Psi}/G_N$ is plotted as a function of redshift.
The posterior obtained from the analysis of LSS data \cite{Ishak:2024jhs} is plotted in green.
The posterior obtained from GWs observations using eq.(\ref{GeffCB0}) is plotted in blue, assuming no-brading, i.e. setting $\alpha_B=0$, and using the posterior for $\eta$ obtained from LSS data analysis \cite{Ishak:2024jhs}. This plot shows the results obtained adopting  the $\alpha_M$ parametrization for the GW-EMW distance ratio. The solid lines correspond to the median, while the shaded bands indicate the $1\sigma$ confidence interval of the posterior.}
\label{GeffCmDESI}
 \end{figure}

For the event GW170817 \cite{LIGOScientific:2017zic} it is possible to obtain a direct estimation of the GW-EMW ratio, since there is a confident association between the GW event and its electromagnetic counterpart, and this type of events are also known as bright sirens. The GW170817 constraints corresponds to the blue data point in fig.(\ref{GeffChiDESI}) and fig.(\ref{GeffCmDESI}). For GW events without an electromagnetic counterpart, also know as dark sirens, it is possible to use statistical methods to perform a joint estimation of cosmological and modified gravity parameters \cite{Chen:2023wpj}. For example for the parametrization 
\be
r_d(z)=\Xi_0+\frac{1-\Xi_0}{(1+z)^n} \label{Xi0}\,,
\ee
the best fit parameters obtained analyzing GWs emitted by black holes and neutron stars binary systems  \cite{Chen:2023wpj} were $\Xi_0 =1.67^{+0.93}_{-0.94}$ and $n=0.8^{+3.59}_{-0.69}$. For the $\alpha_M$ parametrization, corresponding to this expression for the distance ratio
\be
    r_d(z)=\exp \left\{\frac{c_M}{2\Omega_{\Lambda,0}} \ln \frac{1+z}{[\Omega_{m,0}(1+z)^3+\Omega_{\Lambda,0}]^{1/3}} \right\}\,, \label{Cm}
\ee
the best fit value was $c_M=1.5^{+2.2}_{-2.1}$.

LSS observations provide an independent way to test the deviation from GR, and we consider the results obtained analyzing the Dark Energy Spectroscopic Instrument (DESI) data using the parametrization \cite{Ishak:2024jhs}
\bea
\frac{G^{\Psi}_{\rm eff}}{G_N}&=& \mu(a)= \left[1+\mu_0 \frac{\Omega_{\Lambda}(a)}{\Omega_{\Lambda}}\right]\nonumber\,,\\
\frac{G^{\Psi+\Phi}_{\rm eff}}{G_N}&=& \Sigma(a)= \left[1+\Sigma_0 \frac{\Omega_{\Lambda}(a)}{\Omega_{\Lambda}}\right]\label{musigma} \,.
\eea
The best fit values \cite{Ishak:2024jhs}, obtained assuming no scale dependence and a flat $\Lambda$CDM background, are $\{
\mu_0=0.05\pm 0.22 ,\Sigma_0=0.008\pm 0.045\}$ \footnote{DESI+CMB(LoLLiPoP-HiLLiPoP)-nl+DESY3+DESSNY5 analysis}.

The independent estimation of $G^{\Psi}_{\rm eff}$ is obtained from the CR as
\be
\frac{G^{\Psi}_{\rm eff}(z)}{G_ N}=\frac{\bar\eta}{2-\bar\eta}\frac{2}{2-\alpha_B}\left[\frac{d^{\rm GW}_{\Lr}(z)}{  d^{\rm EM}_{\Lr}(z)}\right]^2 \frac{v_{\rm GW}(z)}{v_{\rm GW}(0)}\,. \label{GeffCB0}
\ee
The comparison of the $68\%$ confidence level constraints of $G^{\Psi}_{\rm eff}$ from the LSS and GW observations  is given  in Fig.(\ref{GeffChiDESI}) and Fig.(\ref{GeffCmDESI}), confirming the validity of the consistency relation.
The GW event GW170817 and its electromagnetic counterpart provides a constraint of the effective gravitational constant with an accuracy comparable with LSS constraints, while the analysis of GW events without electromagnetic counterpart are consistent,  but do not have a constraining power comparable to LSS observations.

\begin{figure}
\includegraphics[width=8cm]{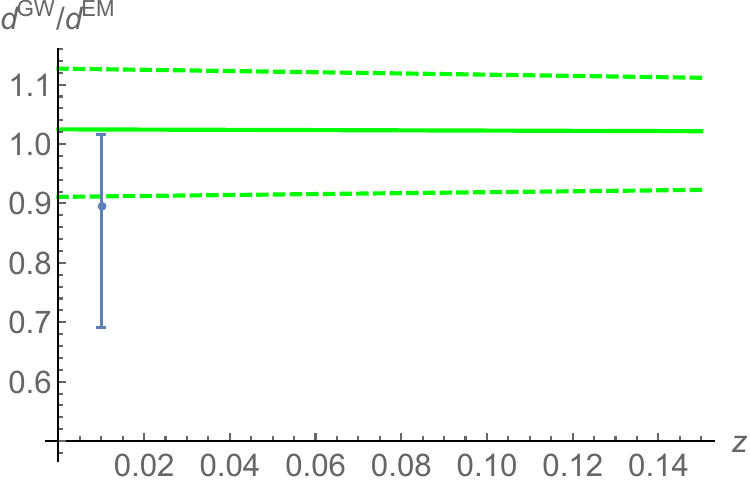}
\caption{The GW-EMW distance ratio implied by non GW observations, obtained using Eq.(\ref{rdGeff}) assuming no-brading no-slip luminal modified gravity theories, is plotted as a function of redshift, using the best fit parameters obtained in \cite{Ishak:2024jhs}. The dashed lines are the $68\%$ confidence  bands. The blue data point corresponds to the bright siren event GW170817 constraint and its $68\%$ confidence interval. The bright siren constraints are consistent with LSS structure observations.
}
\label{rdDESI}
 \end{figure}

\textbf{Comparing LSS and GWs estimations of $r_d$---}    
From the CR we can obtain the  GW-EMW distance ratio implied by LSS observations 
\be
\frac{d^{\rm GW}_{\Lr}(z)}{  d^{\rm EM}_{\Lr}(z)} =\left[\frac{2-\bar\eta}{\bar\eta}\frac{2-\alpha_B}{2}\frac{v_{\rm GW}(0)}{v_{\rm GW}(z)} \frac{G^{\Psi}_{\rm eff}(z)}{G_ N}\right]^{1/2}\,. \label{rdGeff}
\ee
The comparison between the observed GW-EMW distance ratio and the one implied by LSS observations is plotted in Fig.(\ref{rdDESI}), showing that the event GW170817 is consistent with the non GWs observations, and that the constraining power of bright sirens events is comparable to that of LSS observations.

\begin{figure}
\includegraphics[width=8cm]{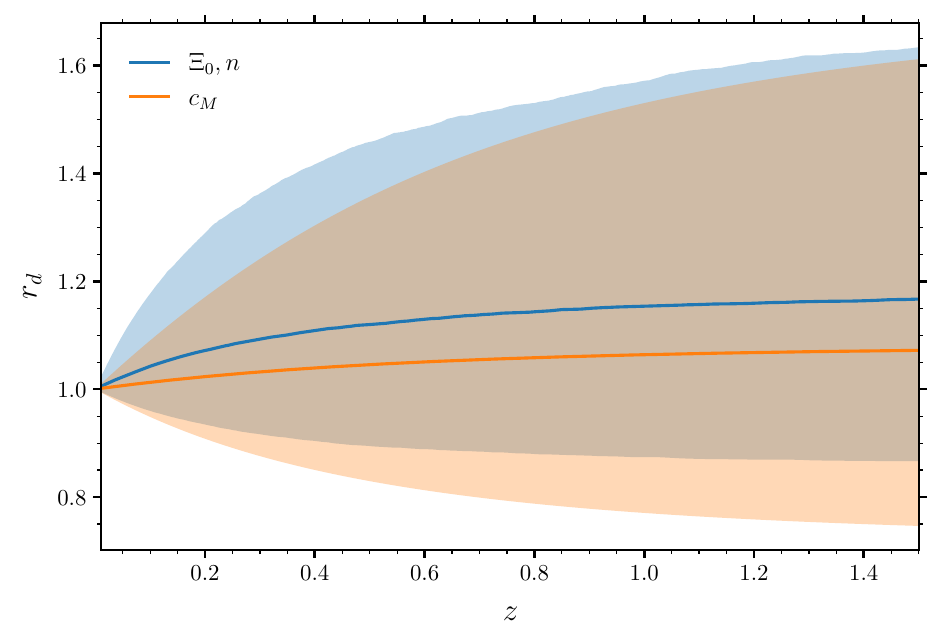}
\caption{GW-EMW distance ratio posterior corresponding to the $\{\Xi_0,n\}$ and $\alpha_M$ parameterizations adopted in \cite{LIGOScientific:2025jau}.}
\label{dist_ratio}
 \end{figure}

\begin{figure}
\includegraphics[width=8cm]{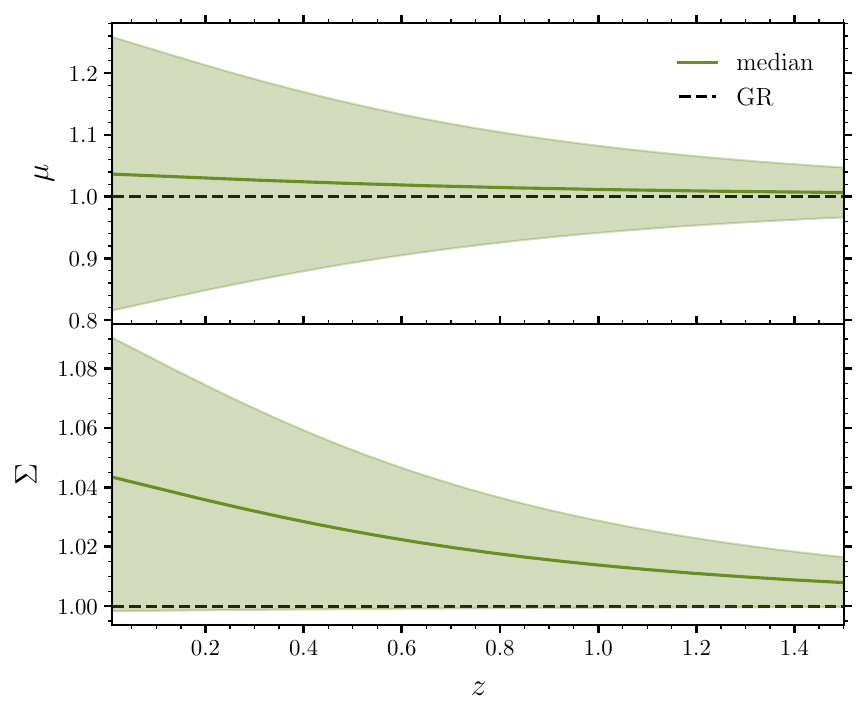}
\caption{Evolution of the modified gravity phenomenological parameters $\mu$  and $\Sigma$  corresponding to the posterior of the  of DESI data \cite{Ishak:2024jhs}. The solid lines correspond to the median, while the shaded bands indicate the $1\sigma$ confidence interval of the posterior. The horizontal dashed line corresponds to the prediction of General Relativity.}
\label{mu_sigma}
 \end{figure}

\begin{figure}
\includegraphics[width=8cm]{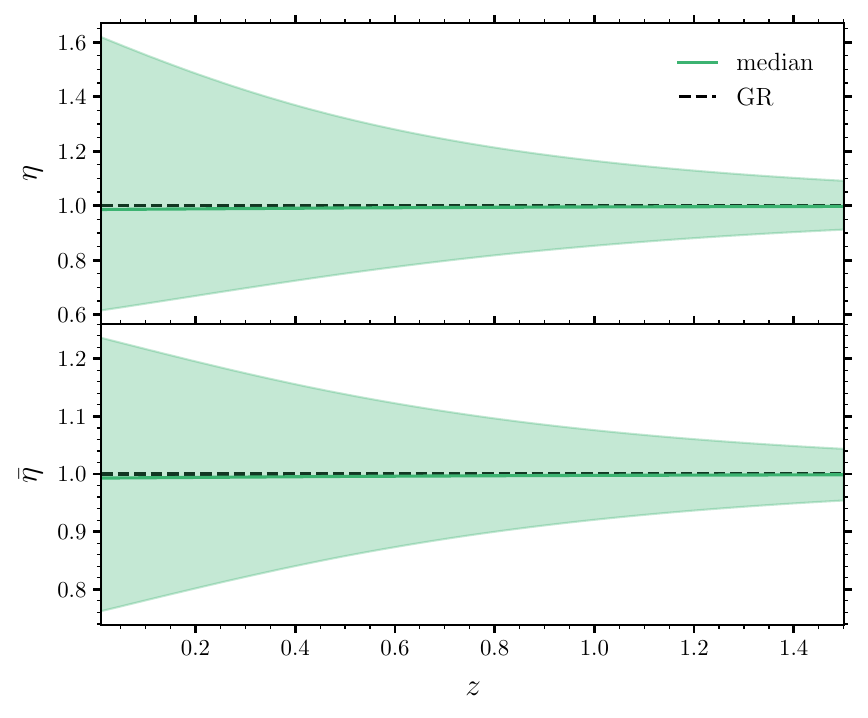}
\caption{Evolution of the gravitational slip parameters $\eta$ and $\bar{\eta}$ corresponding to the $\Sigma$ and $\mu$ posteriors shown in in Fig.(\ref{mu_sigma}), obtained in \cite{Ishak:2024jhs}. The solid lines correspond to the median, while the shaded bands indicate the $1\sigma$ confidence interval of the posterior.
}
\label{slip}
 \end{figure}

\begin{figure}
\includegraphics[width=8cm]{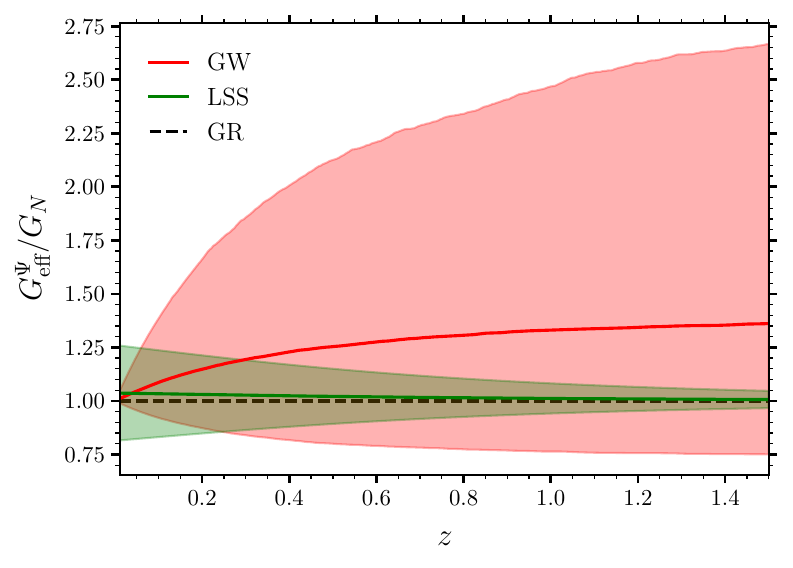}
\caption{The effective gravitational coupling $G_{\text{eff}}^{\Psi}/G_N$ is plotted as a function of redshift.
The posterior obtained from the analysis of LSS data \cite{Ishak:2024jhs} is plotted in green.
The posterior obtained from GWs observations using eq.(\ref{GeffCB0}) is plotted in red, assuming no-slip and no-brading, i.e. setting $\eta=1$ and $\alpha_B=0$. This plot shows the results obtained adopting  the $\{\Xi_0,n\}$ parametrization for the GW-EMW distance ratio. The solid lines correspond to the median, while the shaded bands indicate the $1\sigma$ confidence interval of the posterior.}
\label{Geff_Xi0n_eta1}
 \end{figure}

\begin{figure}
\includegraphics[width=8cm]{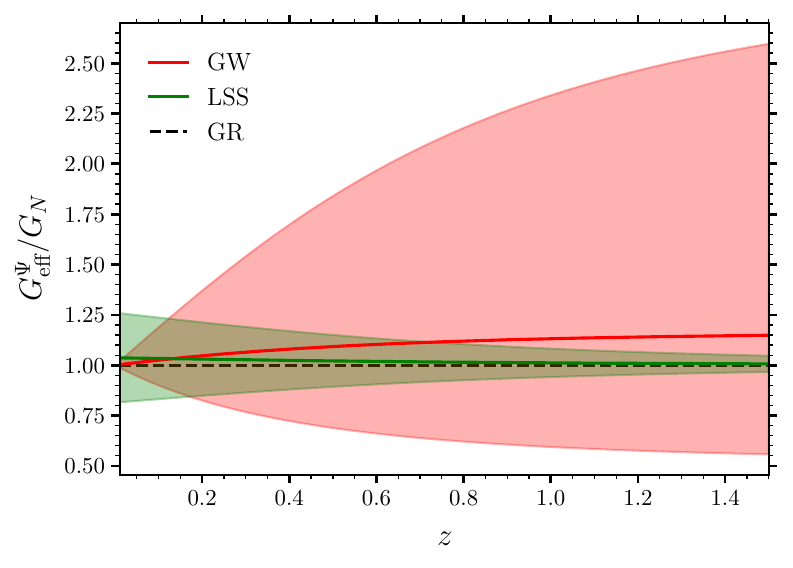}
\caption{The effective gravitational coupling $G_{\text{eff}}^{\Psi}/G_N$ is plotted as a function of redshift.
The posterior obtained from the analysis of LSS data \cite{Ishak:2024jhs} is plotted in green.
The posterior obtained from GWs observations using eq.(\ref{GeffCB0}) is plotted in red, assuming no-slip and no-brading, i.e. setting $\eta=1$ and $\alpha_B=0$. This plot shows the results obtained adopting  the $\alpha_M$ parametrization for the GW-EMW distance ratio. The solid lines correspond to the median, while the shaded bands indicate the $1\sigma$ confidence interval of the posterior.
}
\label{Geff_cM_eta1}
 \end{figure}

\textbf{Conclusions---}
The EFT of dark energy allows to derive   parametrization independent consistency relations between LSS and GWs observation, allowing to test modified gravity and dark energy models by comparing directly independent observations, avoiding the limitations associated to specific choices of parametrization \cite{Linder:2016wqw}, or the difficulty to compare different parametrizations.
We have tested the constant brading consistency condition by comparing the constraints on the effective gravitational constant from LSS and GW observations, confirming its validity at $68\%$ confidence level.
A violation of the CR would imply that the modified gravity effects are due to a theory which cannot be described by the EFT, or that brading is not constant in the the observed redshift range, in which case other CRs \cite{Romano:2025pcs} can be used to account for its time dependency.
Since the GW strain is inversely proportional to the GW luminosity distance, while the apparent magnitude of galaxies is inversely proportional to the square of the electromagnetic luminosity distance, the CRs allow to obtain high redshift estimations of the effective gravitational constant using GW events with an EM counterpart, at distances where large scale structure observations are not available or are not very precise, due to EM selection effects.
The future availability of an increasing number of dark and bright sirens observations will allow to improve the constraints presented in this paper.
Since the left hand side of Eq.(\ref{GeffCB}) depends on LSS observations only, another possible application is to estimate the GWs speed for dark sirens events by combining LSS and GWs luminosity distance observations.

In our analysis we have used the EFT results for the effective gravitational constant and the GW-EMW distance ratio, which are based on the quadratic order action in Eq.(\ref{lgamma}), but higher order effects \cite{Romano:2022jeh,Romano:2023bzn,Romano:2024apw} are expected to introduce a possible scale and polarization dependency of the  speed and the distance ratio. In the future it will be interesting to investigate how this scale dependency is related to that of the effective gravitational constant. We have shown that GWs and LSS observations satisfy the constant brading consistency condition at the current level of experimental uncertainty, but other consistency conditions can be derived assuming different behaviors of $\alpha_B$, so in the future it would be interesting to test other CRs. Nevertheless the advantage of the constant brading CR is that it is also satisfied by GR, and hence it allows to estimate the gravitational coupling from GWs observations also in the standard GR scenario.

In the future it will be interesting to test the CRs with the results of non parametric analyses for both LSS and GWs data, and to include other observations such as  weak lensing and baryonic acoustic oscillations.


\textbf{Acknowledgments---}
This material is based upon work supported by NSF's LIGO Laboratory which is a major facility fully funded by the National Science Foundation.
I thank Johannes Noller, Tessa Baker for their comments, Eric V. Linder, Hsu Wen Chiang and Reginald Christian Bernardo for  interesting discussions, and the Academia Sinica and HCWB for the kind hospitality. We thank Mustapha Ishak for interesting discussions and helping to access the results of the DESI data analysis. This work was supported by the UDEA dedicacion exclusiva program and project No. 2023-63330.

\textbf{DATA AVAILABILITY---}
The data that support the findings of this article are openly available \cite{GWOSC}.

\bibliographystyle{h-physrev4}
\bibliography{mybib}
\end{document}